\documentclass[11pt,twocolumn]{article}
\usepackage{amssymb,amsmath,url}
\usepackage[T1]{fontenc}
\usepackage[utf8]{inputenc}
\usepackage[babel,german=quotes]{csquotes} 
\usepackage{graphicx}
\usepackage{ragged2e}

\begin{document}


\title{Automatic Raga Recognition in Hindustani Classical Music \hfill {\small \bf MUS-17}}
\author{Sanchit Alekh}
\date{
	sanchit.alekh@rwth-aachen.de \\
	\today}

\maketitle

\paragraph{Abstract}
\enquote*{Raga} is the central melodic concept in Hindustani Classical Music. It has a complex structure, often characterized by pathos. In this paper, we describe a technique for Automatic Raga Recognition, based on pitch distributions. We are able to successfully classify ragas with a commendable accuracy on our test dataset.

\section*{Introduction} 
Hindustani Classical Music is a form of music originating in northern India. It was composed between $1500-900 B.C$\cite{Overview}, making it widely regarded as one of the oldest extant music systems in the world \cite{Old}. One of the most striking features of hindustani music is its imaginative and improvised nature, i.e there is neither a written script while performing, nor fixed compositions that the artist has to adhere to. In effect, the entire performance is an extempore, and the artist plays the role of the singer, composer and the conductor at the same time.
At the center of Hindustani Music is a melodic concept called the \enquote*{Raga}. According to Rao et al.\cite{Overview}, Raga is a continuum with scale and tune as its extremes. Broadly speaking, it can be termed as a melodic mode or tonal matrix possessing a rigid and specific individual identity, yet bearing immense potential for infinite improvisatory possibilities. The raga serves as a basic grammar for composition and improvisation in Indian music.

\subsection*{Raga Recognition: The Problem}
The use of statistical and probabilistic tools in musicology is not new\cite{Temperley,Beran}. A strong theoretical grounding in Computational Musicology has sparked interest in the subject of Automatic Raga Recognition (ARR) in recent years. In a crude form, \enquote*{Raga Recognition} refers to techniques for identifying the raga in which an artist performs his rendition. Due to the complex nature of a raga, as well as nuanced differences between several ragas, this is not a trivial problem. Moreover, in Hindustani Music, the tonic is never fixed, i.e. an artist can perform the same raga in different tonic scales on different occasions. For this reason, ARR is often also accompanied with, or preceded by tonic identification.

\section*{Efforts in Raga Recognition}
The current state-of-the-art approaches in ARR take two different paths to tackle the same problem. The first approach, developed by Chordia et al. \cite{Chordia}, is based on pitch distributions. Here, the authors systematically compare the accuracy of several methods using tonal features, combined with nearest-neighbour techniques and bayesian classifiers. They calculate the pitch distributions for different tonic pitches and try to find the one that gives the best match to a database of samples. In the second approach, developed by Gulati et al.\cite{Gulati}, the authors use a vector-space model of the melodic phrases, which are extracted in an unsupervised manner. After representing audio recordings in a suitable vector-space, they employ a number of classification strategies to build a predictive model for raga recognition such as support vector machines,stochastic gradient descent learning, logistic regression and random forests. 

Both these approaches have proven themselves to perform really well on various datasets, outperforming one another in some tests. The maximum accuracy that has currently been reached is $91.7\%$. The approach proposed by Chordia et al. formed the basis of my studies during the course of this seminar, and this is the method that I present in this paper.

\section*{Raga Recognition based on Pitch Distributions}
Chordia et al. \cite{Chordia} proposed a method for simultaneous recognition of tonic and raga based on pitch distributions. They model the pitch distribution of a frame non-parametrically, which results in a robust feature for classification. The novelty of this work is the use of high-dimensional pitch representations (HPDs), namely the Fine-Grained Pitch Distribution (FPD) and Kernel-Density Pitch Distribution (KPD), which overcome several limitations of the simple histogram approach. As illustrated in \textit{Figure 1}, the approach is broadly divided into 4 stages: 
\begin{enumerate}
	\item Pitch Tracking
	\item Tonal Feature Extraction
	\item Tonic Estimation
	\item Raga Recognition
\end{enumerate} 

In the next sections, I will describe these stages one-by-one.

\begin{figure}[ht]
	\centering
	\includegraphics[width=8.2cm]{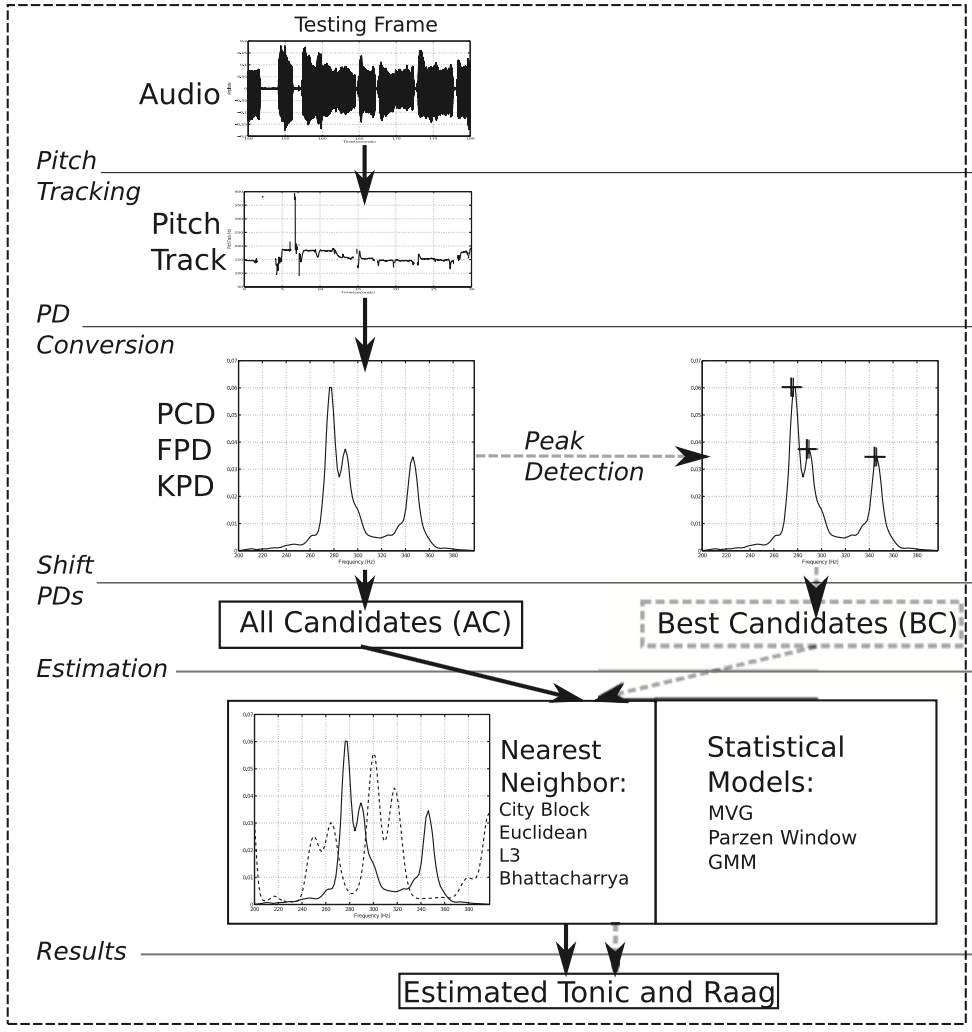}
	\caption{Stages in Simultaneous Tonic and Raga Recognition. \textit{Source: Chordia et al.}\cite{Chordia}}
	\label{fig:stages}
\end{figure}

\subsection*{1. Pitch Recognition}
For tracking the pitch of the input audio excerpt, the authors use the Sawtooth-Waveform-Inspired Pitch Estimator (SWIPE') \cite{Camacho}. The SWIPE algorithm estimates the pitch as the fundamental frequency of the sawtooth waveform whose spectrum best matches the spectrum of the input signal. The pitch of the audio input is estimated every $10 ms$, and the estimate is constrained to the range $73.4-587.2 Hz$ using a resolution of $48$ steps per octave. Apart from the pitch estimate, the SWIPE algorithm also returns an estimate of pitch strength, which is a value between $0$ and $1$. The frequencies for which the pitch strength is less than $0.2$ are deemed unreliable and replaced with \textit{\enquote*{NaN}}.

\subsection*{2. Tonal Feature Extraction}
The pitch tracks obtained in Stage 1 are used to compute pitch distributions (PDs), which are the fundamental tonal features for the raga-recognition task, along with tonic frequency. As mentioned earlier, the approach uses three different kinds of histograms, namely the 12-dimensional PCD (Pitch Class Distribution), FPD and KPD, out of which the latter two are HPDs. This is done because a 12-dimensional categorization is inadequate for capturing ornamentations such as vibrato (andolan) and portamento (meend), and the blending of notes (swaras) during a performance. Some of these embellishments are essential and discriminating features of a raga, and therefore, it is important that they are identified correctly. HPDs also aid us in tracking the microtonal pitch information, which is essential in identifying several ragas, especially ragas from the \textit{Kanhada} and the \textit{Malhar} family. The end effect is the precise capturing of a raga-specific characteristics, without modeling sequential information. \textit{Figure 2} illustrates the different pitch distributions calculated during this stage.

\begin{figure}[ht]
 	\centering
 	\includegraphics[width=8.2cm]{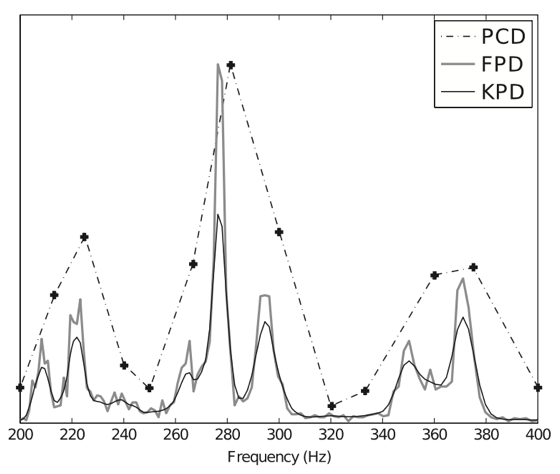}
 	\caption{Tonal Feature Extraction from Raga Audio. \textit{Source: Chordia et al.}\cite{Chordia}}
 	\label{fig:features}
\end{figure}

Before the pitch distributions are calculated, all pitches are mapped into one octave by dividing or multiplying each pitch frequency by $2^{k}$ for the value of $k$, which places it within the desired octave. The PCD is then calculated by mapping each of the pitch estimates to one of the 12 chromatic pitch classes. The PCD is then normalized to add up to 1, to make it independent of frame duration.

The HPDs, i.e. FPD and KPD, are much more continuous than the PCD. They are calculated in a similar fashion. In the FPD, for example, instead of 12 bins, 120 or 240 bins are used, for a bin width of 10 cents and 5 cents respectively. The KPD, on the other hand, is essentially a continuous pitch distribution, and is calculated using \textit{Kernel density estimation (KDE)}, also known as the \textit{Parzen Window Method}. Instead of assigning a pitch value, KDE centers a window function, such as Gaussian curve on the pitch value. The overall pitch density is the sum of all such gaussian curves, i.e. a convolution of the different impulses on the pitch values. The gaussian kernel density function is given by the following equation, 

\begin{equation}
	f_h(x) = \frac{1}{nh} \sum_{i=1}^{n} \frac{1}{\sqrt{2 \pi}} e^-{\frac{(x-x_i)^2}{2h^2}}
\end{equation}
where, \\
$h$ is the width of the Gaussian kernel \\
$x_i$ is the value of the i-th pitch, and \\
$n$ is the total number of pitch values.

\subsection*{3. Tonic Estimation}
As mentioned earlier, tonic estimation is an important pre-requisite for raga recognition. To motivate this, let us take an example. In the scale of C, \textit{Raga Malkauns} ascends like $(C, D^{\#}, F, G^{\#}, A^{\#}, C)$. However, if the tonic is placed on $D^{\#}$, the same set of notes identify \textit{Raga Durga}, which is an entirely different raga from a different family. 
Tonic Estimation is performed using two approaches, and both of them are based on calculating the PDs for different candidates and finding the one with the best match to the database. In the PD curve, the \textit{x-axis} represents the number of cents above the tonic frequency, therefore changing the tonic also changes the PD curve. The difference between the two methods is the manner in which these candidates are selected. The \textit{All-Candidates}(AC) approach uses brute force, and all \textit{120 (or 240)} frequencies are used to calculate the PD in a circular fashion. This leads to \textit{120 (or 240)} different hypotheses. In the \textit{Best-Candidates}(BC) approach, the PD is only created for frequencies which appear as peaks in the HPD. By considering the $7$ highest peaks, computation is greatly reduced.

In both these approaches, the hypotheses are matched to all samples in the database and the nearest neighbour(NN) is found. The NN with the minimum overall distance is taken as the tonic.

\subsection*{4. Raga Recognition} 
The raga is simultaneously recognized during tonic estimation. It is either the label of the overall nearest training sample, or the raga category which yields the maximum posterior probability. For the NN approach, several distance measures, such as city block, Euclidean and Bhattacharya distance are used. The most popular method for comparing probability densities is the \textit{Bhattacharya Distance}, and it also gives us the best results in our experiments. It is computed as,

\begin{equation}
	D_B(p,q) = - \ln \bigg(\sum_{i=1}^{n} \sqrt{p_i q_i}\bigg)
\end{equation}
where $p = (p_1, p_2, \dots p_n)$ and $q = (q_1, q_2, \dots q_n)$

In the statistical approach, we use the Bayes' Rule\cite{Duda} to calculate the conditional probability

\begin{equation}
	P(raga_i\mid x) = \frac{P(x\mid raga_i) P(raga_i)}{\sum_{j} P(x\mid raga_j) P(raga_j) }
\end{equation}
where $x$ is one of the test PDs. The posterior probability $P(x\mid raga_i)$ is estimated empirically using parametric density models such as multi-variate Gaussian (MVG) and Gaussian Mixture Models (GMM).

\section*{Results and Conclusion}
For the experiments, the authors used the database \textit{GTraagDB}, available at \textit{paragchordia.com/data/GTraagDB}. The database has $127$ samples from $31$ different ragas. The tests are performed by modifying several configurable parameters, such as precision (for tonic estimation), granularity of the KPD, kernel width, distance algorithm and all/best candidates approach.

For tonic estimation as well as raga recognition, KPD with 5-cent granularity, combined with the NN classifier with the Bhattacharya Distance (NNB), with the AC approach and a kernel-width of $0.1 Hz$ gives us the best results. In case of tonic estimation, the minimum error rate for 15-cent precision was obtained as $4.92 \%$. For raga estimation, the same configuration yielded a minimum error rate of $8.5 \%$. \textit{Figure 3} illustrates the comparison between the average error rates for raga recognition using different classifiers.

\begin{figure}[ht]
	\centering
	\includegraphics[width=8.2cm]{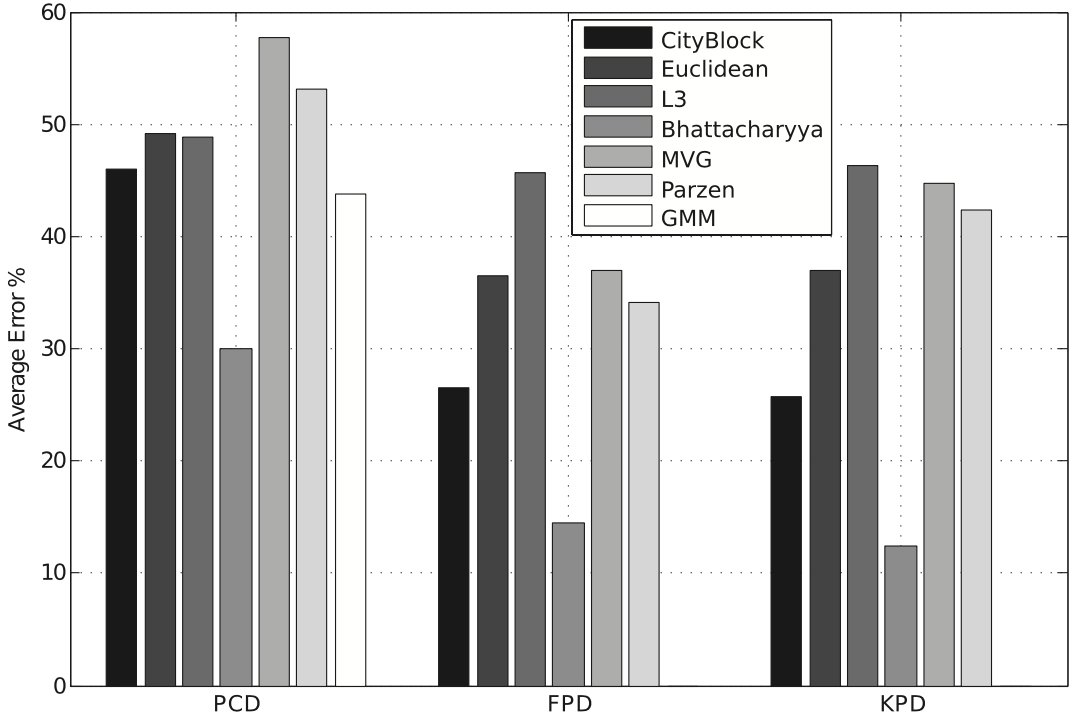}
	\caption{Average Raga Error Rates for AC Method. \textit{Source: Chordia et al. \cite{Chordia}}}
	\label{fig:results}
\end{figure}

The difference between the results obtained using NNB and other approaches is quite startling. The next best classifier for PCD has an error rate of $16\%$ higher. It is also observed that the statistical approach performs almost as worse as the other classifiers with an exception of NNB. Additional results from the experiments can be found in \textit{Appendix 1}.

In conclusion, this approach is successfully able to recognise ragas with a good accuracy. This provides evidence that melodic estimation is possible and effective in a complex musical genre with continuous pitch movements and diversity of scale types. The experiments also concretize the intuition that richer, more fine-grained distributions should perform better than PCD. An analysis of misclassified results show that the automatic approach confuses between some typical ragas, such as \textit{Desh} and \textit{Khamaj}, \textit{Asavari} and \textit{Darbari} etc., which are sometimes difficult to distinguish even for seasoned listeners. The results from these experiments also give us an intuition that capturing temporal information, possibly using \textit{Hidden Markov Models} can make the classification even better. This can be used for future improvements in ARR.

\onecolumn
\newpage

\begin{appendix}
  \section*{Appendix 1}
  \begin{figure}[ht]
  	\centering
  	\includegraphics[width=9.5cm]{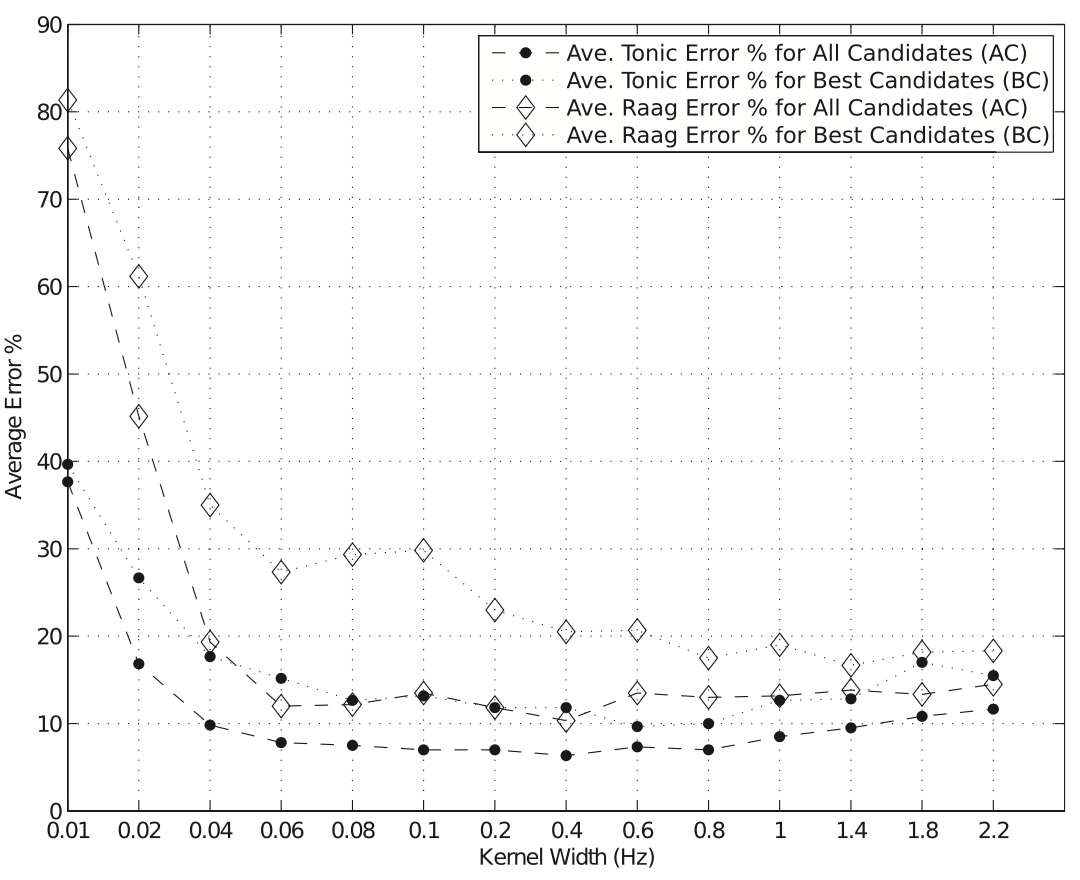}
  	\caption{Average Error Rates for different Kernel Widths for NNB}
  	\label{fig:kernel}
  \end{figure}

  \begin{figure}[ht]
  	\centering
  	\includegraphics[width=9.5cm]{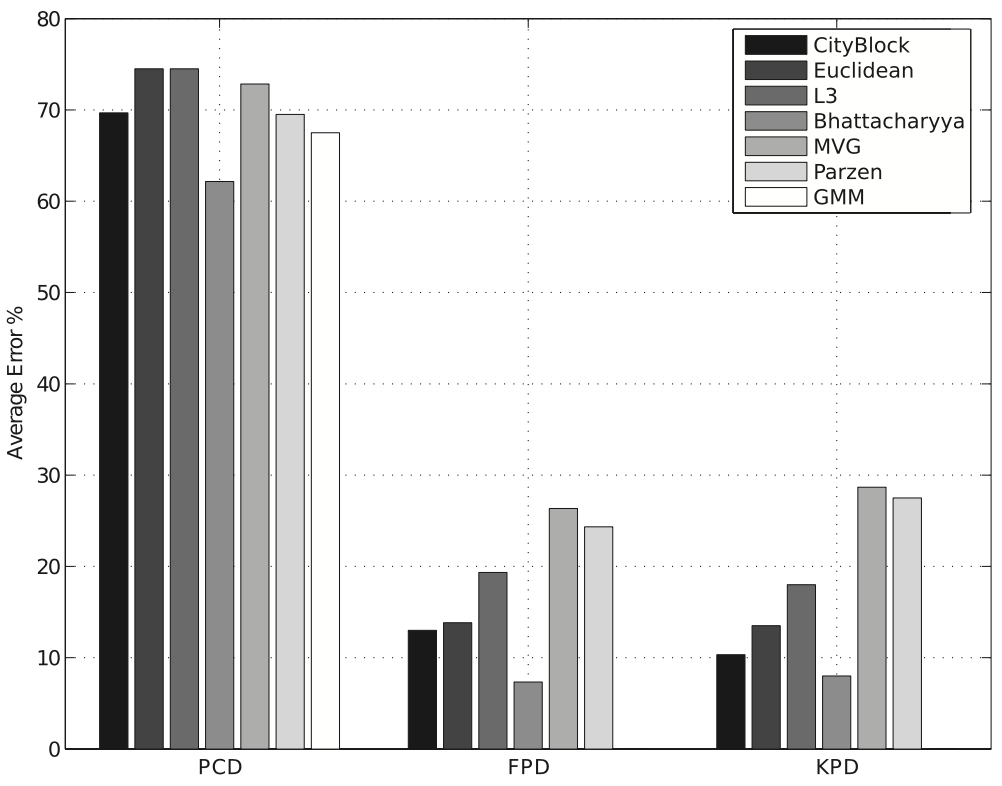}
  	\caption{Average Tonic Error Rates for the AC Method}
  	\label{fig:tonic}
  \end{figure}

  \begin{figure}[ht]
  	\centering
  	\includegraphics[width=9.5cm]{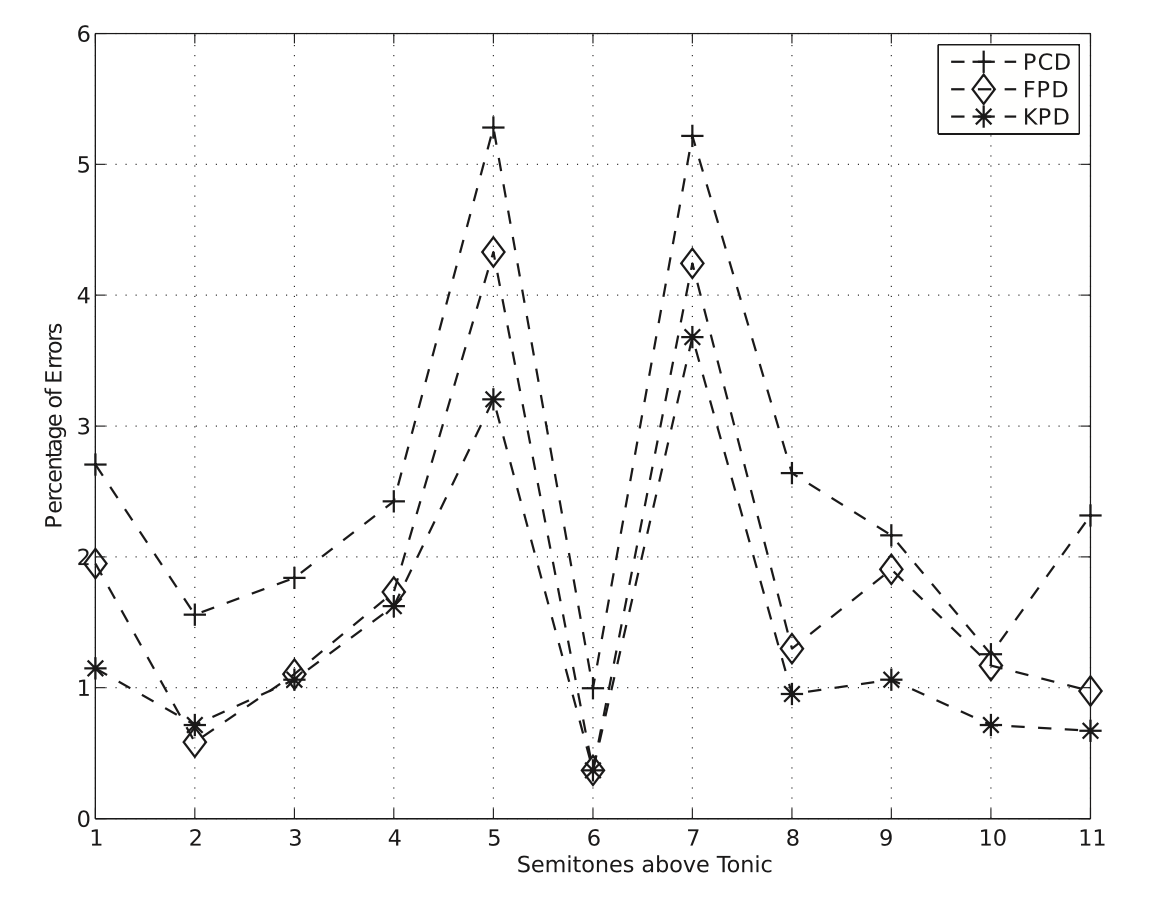}
  	\caption{Average Tonic Error Rates with respect to different Semitones}
  	\label{fig:semitones}
  \end{figure}

  \begin{figure}[ht]
  	\centering
  	\includegraphics[width=\textwidth]{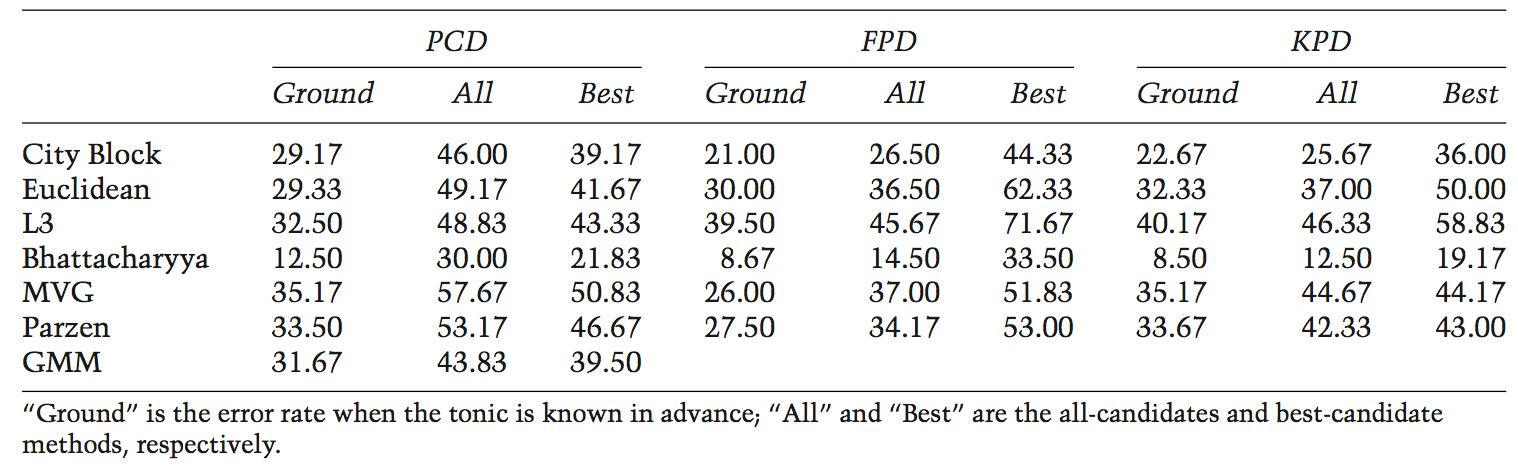}
  	\caption{Average Raga Error Rates for the different Features}
  	\label{fig:average}
  \end{figure}
\end{appendix}


\begin{thebibliography}{99}

\bibitem{Overview}
S.~Rao and P.~Rao.
{\sl An Overview of Hindustani Music in the context of Computational Musicology.}
Journal of New Music Research, vol. 43, no. 1, pp.~24–33, 2014.

\bibitem{Old}
Slawek, Stephen.
{\sl Improvisation in Hindustani Instrumental Music.}
In the course of performance: Studies in the world of musical improvisation (1998): 335.

\bibitem{Temperley}
D.~Temperley.
{\sl Music and Probability.}
The MIT Press, 2007.

\bibitem{Beran}
J.~Beran.
{\sl Statistics in Musicology.}
CRC Press, 2003.

\bibitem{Chordia}
P.~Chordia and S.~Sentürk.
{\sl Joint Recognition of Raag and Tonic in North Indian Music.}
Computer Music Journal, vol. 37, no. 3, pp. 82–98, 2013.

\bibitem{Gulati}
S.~Gulati, J.~Serra, V.~Ishwar, S.~Sentürk, and X.~Serra.
{\sl Phrase-based Raga Recognition using Vector Space Modeling.}
IEEE International Conference on Acoustics, Speech and Signal Processing (ICASSP), IEEE, 2016, pp.~ 66–70.

\bibitem{Camacho}
Camacho, Arturo.
{\sl SWIPE: A Sawtooth Waveform Inspired Pitch Estimator for Speech and Music.}
University of Florida, Gainesville, 2007.

\bibitem{Duda}
Duda,~Richard O., Hart,~Peter E. and Stork,~David G. 
{\sl Pattern classification}. 
John Wiley \& Sons, 2012.

\end{thebibliography}
\end{document}